# Phonon and electronic structures and resistance of layered electride $Ca_2N$: DFT calculations

B. N. Mavrin, M.E. Perminova and Yu.E.Lozovik

*Institute of Spectroscopy of RAS, Moscow Troitsk 142190, Russia*

**Abstract**

The phonon and electronic properties, the Eliashberg function and the temperature dependence of resistance of electride $Ca_2N$ are investigated by the DFT-LDA plane-wave method. The phonon dispersion, the partial phonon density of states and the atomic eigenvectors of zero-center phonons are studied. The electronic band dispersion and partial density of states conclude that $Ca_2N$ is a metal and the Ca 3p, 4s and N 2p orbitals are hybridized. For the analysis of an electron - phonon interaction (EPI) and its contribution to resistance the Eliashberg function was calculated and a temperature dependence of resistance caused EPI was found. The present results are in good agreement with experiment data.

## 1. Introduction

Recently, more and more attention has been attracted to layered electrides in bulk and monolayer form in which playing anionic role electrons form two-dimensional planes separated from positively charged layers of ions [1-6]. In view of their promising properties such as high electrical conductivity, low work function, and significant catalytic activity in their ideal form, electrides are drawing much attention from the research community.

Sub-nitride $Ca_2N$ belongs to this new class of layered-structure electrides with the two-dimensional delocalized layers of electrons [7]. The unit cell of $Ca_2N$ contains more electrons than it is expected from the simple electron counting rules. It is supposed that excess electrons are localized between positively charged layers $[Ca_2N]^+$. The possible applications of this material in electronics were discussed recently [7-9]. The physical properties of $Ca_2N$ were studied experimentally by photoelectron spectroscopy [7,9], optical reflectance spectroscopy [7], electrical conductivity [7,10], by study of magnetic susceptibility [10], magnetoresistance [7] and theoretically [7-9,11,12]. It is found that single crystal $Ca_2N$ exhibits metallic transport with resistivity, which is smaller than that of pure Ca metal [7]. The temperature dependence of resistivity indicated that the electron-electron interaction could be much stronger than the electron-phonon interaction even in the high-temperature region [7]. A high electron mobility and low work function are revealed from measurements [7].

The phonon dispersion was calculated only for isolated layer Ca$_2$N [12]. The electronic structures were studied earlier in rhombohedral [7,11] and hexagonal [8,12] settings, using projected augmented plane-wave method [7,11,12] or localized spherical method [11]. The resistance from first-principles can be found only with norm-conserving pseudopotentials in plane-wave basis.

In the present work we considered semi-core electrons for the Ca atom at calculations of electronic and phonon properties of Ca$_2$N. We have computed the phonon and electronic band dispersion in the Brillouin zone (BZ) of crystal, the partial phonon and electronic density of states and we have found the eigenvectors for zone-center optical phonons. For an estimation of an electron - phonon interaction (EPI) and its contribution to resistance the Eliashberg function was calculated and a temperature dependence of resistance caused EPI was found.

## 2. Computational method

To calculate the phonon properties we have used the ab initio plane-wave program ABINIT in local density approximation with the Hartwigsen-Goedecker-Hutter (HGH) relativistic pseudopotentials [13], Monkhorst-Pack (MP) mesh [14] of 4x4x4, kinetic energy cutoff of 40 Ha. The electron configurations $3s^23p^64s^2$ and $2s^22p^3$ were used for the Ca and N atoms, respectively. The MP mesh was increased to 8x8x8 in the study of electronic and structural properties and to 16x16x16 in the EPI calculations.

Ca$_2$N crystallizes in the rhombohedral structure (space group $R\bar{3}m(D_{3d}^5)$, Z=1 (Fig. 1(a)) [15-17] forming the Ca-N-Ca layers (Fig. 1(b)). The Ca and N atoms occupy the $C_{3v}$ and $D_{3d}$ sites in the unit cell, respectively. After the structural optimization the lattice parameters decreased by about 1 %.

## 3. Results and discussion

### 3.1. Phonon properties

The unit cell of Ca$_2$N contains 3 atoms resulting in 9 phonon branches with Γ-point symmetries 2A$_{2u}$ + A$_{1g}$ + E$_g$ +2E$_u$. Of these 1A$_{2u}$ + 1E$_u$ comprise the acoustic modes; 1A$_{2u}$ + 1E$_u$ phonons are infrared active and A$_{1g}$ + E$_g$ are Raman active modes.

The frequencies of phonon branches have been calculated within density perturbation theory [18]. Phonon dispersion relations along the high symmetry directions in the BZ (Fig. 1(c)) are presented in Fig. 2(a). The dynamical matrices have been computed for 200 $\vec{q}$ points along T-Γ-F-L-Γ path. There is no an energy gap between acoustic and optic branches. In Γ-point the frequencies of optic modes at 176, 322, 327 and 415 cm$^{-1}$ are assigned to E$_g$, A$_{1g}$, E$_u$ and A$_{2u}$ symmetries, respectively. In comparison with calculations of phonons in a monolayer Ca$_2$N [12] splitting between frequencies of $E_u$ and $A_{2u}$ modes significantly increased. The atoms in the $E_g$ and $A_{1g}$ modes move in out-plane direction, while $E_u$ and $A_{2u}$ symmetries are in-plane modes (Fig. 2(c)). In $E_g$ and $A_{1g}$ modes only the Ca atoms move in opposite directions, while in $E_u$ and $A_{2u}$ modes two atoms Ca and the N atom move towards.

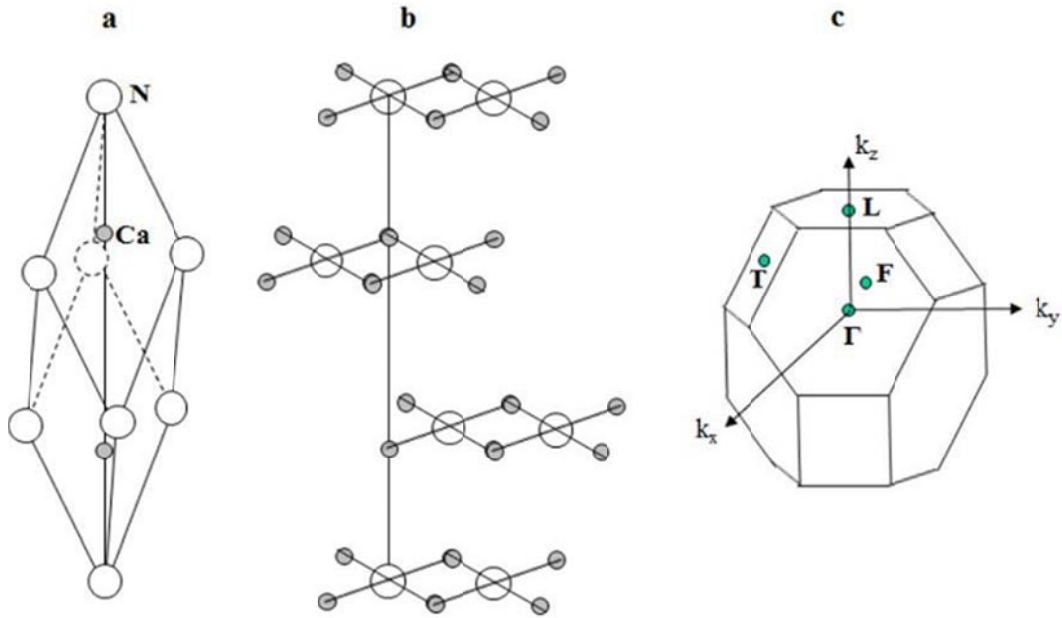

**Fig. 1.** Rhombohedral cell of $Ca_2N$ (a), a layer structure of $Ca_2N$ (b) and Brillouin zone (c).

Phonon density of states (PDOS) is shown in Fig. 2(b). For the PDOS calculation we have applied the tetrahedron method on a mesh of 20x20x20 points in the irreducible BZ using a Fourier interpolation technique provided the dynamical matrices at the other points of the BZ. If the main contribution of the N atoms in PDOS locates above 285 cm$^{-1}$, the Ca atom gives a contribution at lower frequencies except for acoustic modes in which a participation of both atoms is noticeable.

### 1.1. Electronic properties

Electron dispersion relations along the high symmetry directions of the BP (Fig. 1(c)) and partial electronic densities (DOS) are shown in Fig. 3. Direct comparison with results of the previous calculations of the band dispersion [7,11] is difficult as in them the main attention was paid to dispersion in the plane of layers. As in [7,11], there is no band crossing the Fermi level in the Γ-L direction perpendicular to layers Ca-N-Ca (Fig. 3(a)). The "interstitial band", that supposedly [7] is related to two-dimensional delocalized layers of electrons between positively charged layers $[Ca_2N]^+$, crosses the Fermi level in the layer directions Γ-T, Γ-F and F-L.

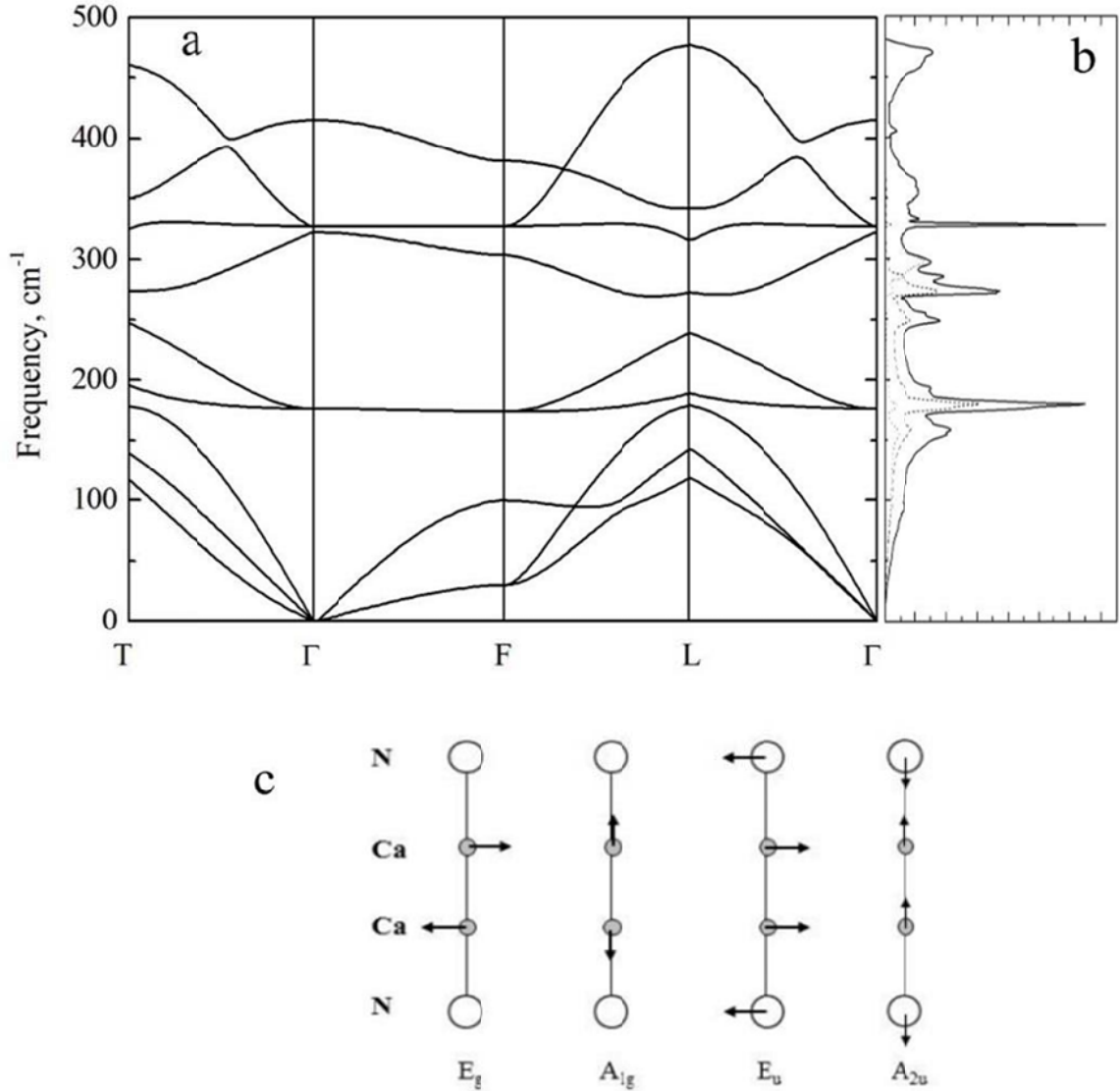

**Fig. 2**. Phonon dispersion (a), partial phonon DOS (b) of $Ca_2N$ (solid line), Ca (dashed line) and N (dotted line) and eigenvectors (c) of zero-center optical modes.

The valence bands near the Fermi level are mainly composed of N 2p states mixed with the Ca 4s and Ca 4 p states (Fig. 3(b)). The DOS at the Fermi level is finite, that is $Ca_2N$ is a metal, and the DOS contains almost equal contributions of the N 2p and Ca 4p states. As distinct from the DOS in the localized spherical approximation [5], the N 2p states show 3 maxima (-1.54, -2.0 and -2.7 eV), the Ca 4s states 2 maxima (-2.0 and -2.7 eV) and the Ca 3p states 3 maxima (-1.6, -1.9 and -2.65 eV). The partial electronic densities of the N 2p and Ca 3p, 4s states are overlapped and these states are hybridized.

*3.3. Eliashberg function and electrical resistance*

The role of different phonon branches in electron-phonon interactions (EPI) can be characterized by the spectral Eliashberg function. The Eliashberg function is a product of the squared effective EPI ($\alpha^2$) and the phonon density of states $F(\omega)$ [19]. The function $\alpha^2 F(\omega)$ is given by [20]:

$$\alpha^2 F(\omega) = \frac{1}{N(\epsilon_F)} \sum_{\vec{k},\vec{q},v,n,m} \delta(\epsilon_{\vec{k}}^n) \delta(\epsilon_{\vec{k}+\vec{q}}^m) \left|g_{\vec{k},\vec{k}+\vec{q}}^{v,n,m}\right|^2 \delta(\omega - \omega_{\vec{q}}^v),$$

where $g_{\vec{k},\vec{k}+\vec{q}}^{v,n,m}$ is the electron-phonon matrix element of interaction between the $\omega_{\vec{q}}^v$ phonon and the electronic states $\epsilon_{\vec{k}}^n$ and $\epsilon_{\vec{k}+\vec{q}}^m$. The calculations of $\alpha^2 F(\omega)$ were performed by the method [21] which is implemented in ABINIT.

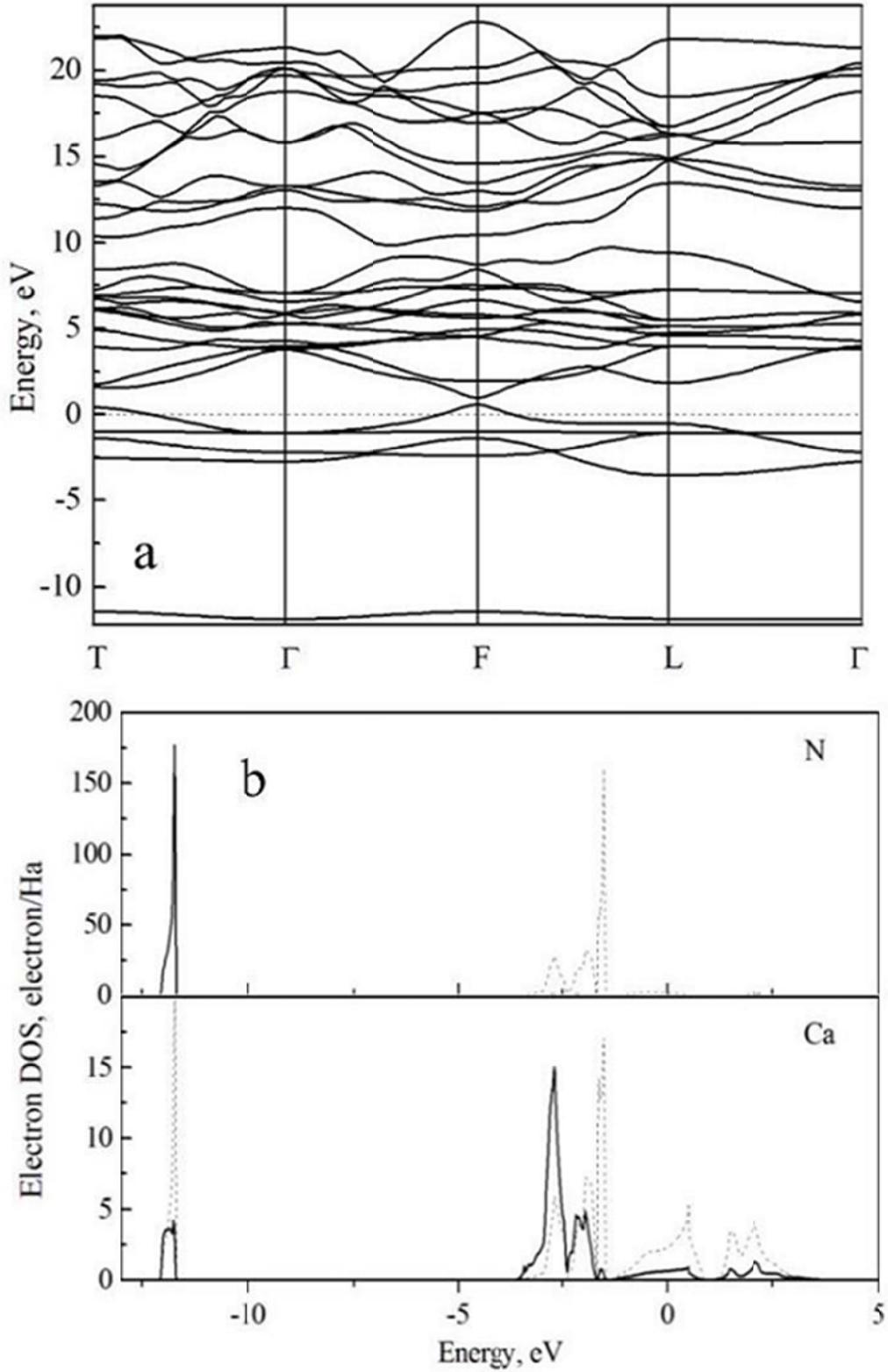

**Fig**. **3**. Electronic band dispersion (a) and partial DOS (b) of s states (solid line) and p states (dashed line) for N and Ca atoms. The Fermi energy is set at 0 eV.

The calculated Eliashberg function for $Ca_2N$ is shown in Fig. 4(a). The main peaks in phonon DOS (Fig. 2(b)) are seen also in Fig. 4(a), confirming a contribution of almost all phonons to the Eliashberg function in $Ca_2N$.

The electrical resistance of a crystal without impurities is determined by the electron-electron and electron-phonon scatterings [20]. Here we calculate the resistance only due to electron-phonon interactions. In this case the temperature dependence of electrical resistivity can be calculated in the lowest-order variational approximation [21]:

$$\rho(T) = \frac{\pi \Omega_{cell} k_B T}{N(\epsilon_F)\langle v_x^2 \rangle} \int_0^\infty \frac{d\omega}{\omega} \frac{x^2}{\sinh^2 x} \alpha^2 F(\omega),$$

where $\alpha^2 F(\omega)$ is Eliashberg's function, $N(\epsilon_F)$ the density of electronic states at the Fermi level and $\langle v_x^2 \rangle$ an average square of the Fermi velocity for electrons and $x = \hbar\omega/2k_B T$. The calculated resistance $\rho$ of $Ca_2N$ is strongly anisotropic (Fig. 4(b)). If the in-plane resistance $\rho_x$ is equal to $0.9 \cdot 10^{-6}$ Ω m at 300 K, then $\rho_z$ is much greater. The small value of $\rho_x$ is an agreement with idea of presence of a two-dimensional layer of electrons between positively charged layers $[Ca_2N]^+$ [7,8]. According to measurements of resistance at 300 K [7], $\rho_x$ (exp) = $2.8 \cdot 10^{-4}$ Ω m. The temperature course of the measured resistance allowed authors [7] to come to conclusion that the contribution of electron-electron scattering to resistance exceeds a the conclusion of authors [7]. On the other hand, the measured resistance in a polycrystalline $Ca_2N$ at 300 K ($\sim 1.0 \cdot 10^{-2}$ Ω m [7]) exceeds also the value calculated perpendicular to layers that can be due to contribution of borders between crystallites at measurements and the primary contribution of the electron-electron scattering.

## 2. Conclusion

We have presented a first-principles study of phonon and electronic structures and resistance of electride $Ca_2N$ at the LDA level taking into account semi-core electrons for the Ca atom and using the plane-wave pseudopotential method. The phonon properties such as the phonon dispersion, the partial phonon density of states and the eigenvectors of zone-center optic modes are investigated. The electronic properties and hybridization of the Ca 3s, 3p and N 3p orbitals are discussed by their density of states. The analysis of electrical resistance concludes the primary contribution of electron-electron processes. The present calculation results are in good agreement with available experiments.

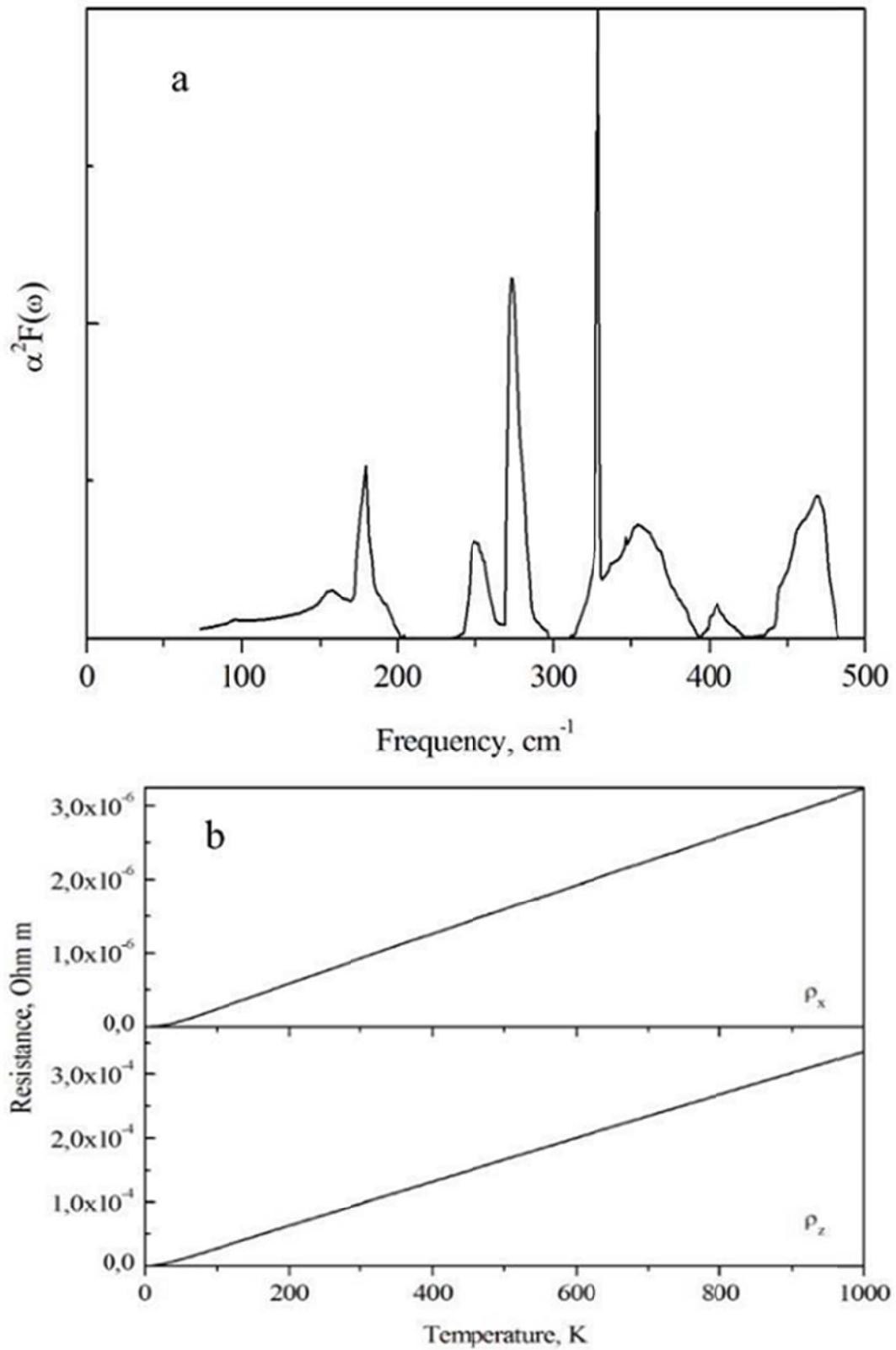

**Fig**. 4. Eliashberg function (a) and resistance (b) along layers ($\rho_x$) and perpendicularly to layers ($\rho_z$).

**Acknowledgments**

We would like to thank Dr. T.A. Ivanova for assistance. The use of facilities of the Joint Supercomputer Center of RAS is greatly appreciated. The work was supported by RFBR grant 17-02-01134. Yu.E.L was supported by the Program of Basic Research of HSE.